\documentclass[final,3p,times]{elsarticle}
\usepackage[utf8]{inputenc}
\usepackage{graphicx}
\usepackage{bm}%
\usepackage{siunitx}
\usepackage{textcomp}
\usepackage{lineno}
\usepackage{xcolor}
\usepackage{subfigure}
\usepackage{amsmath}
\usepackage{textcomp}
\usepackage{multicol}
\usepackage{multirow}
\usepackage{rotating}
\usepackage{hyperref}
\biboptions{sort&compress}
\journal{XXXX}

\begin{document}

\begin{frontmatter}


\title{An electromagnetic physics constructor for low energy polarised X-/gamma ray transport in Geant4}



\author{Jeremy~M.~C.~Brown$^{a,b}$, and Matthew~R.~Dimmock$^{c}$}
\address{$^{a}$Department of Radiation Science and Technology, Delft University of Technology, The Netherlands \\ $^{b}$Centre for Medical Radiation Physics, University of Wollongong, Wollongong, Australia \\ $^{c}$Department of Medical Imaging and Radiation Sciences, Monash University, Melbourne, Australia}

\begin{abstract}
The production, application, and/or measurement of polarised X-/gamma rays are key to the fields of synchrotron science and X-/gamma-ray astronomy. The design, development and optimisation of experimental equipment utilised in these fields typically relies on the use of Monte Carlo radiation transport modelling toolkits such as Geant4. In this work the Geant4 ``G4LowEPPhysics" electromagnetic physics constructor has been reconfigured to offer a ``best set" of electromagnetic physics models for studies exploring the transport of low energy polarised X-/gamma rays. An overview of the physics models implemented in ``G4LowEPPhysics", and it's experimental validation against Compton X-ray polarimetry measurements of the BL38B1 beamline at the SPring-8 synchrotron (Sayo, Japan) is reported. ``G4LowEPPhysics" is shown to be able to reproduce the experimental results obtained at the BL38B1 beamline (SPring-8) to within a level of accuracy on the same order as Geant4's X-/gamma ray interaction cross-sectional data uncertainty (approximately $\pm$ 5 \%).

\end{abstract}


\end{frontmatter}


\section{Introduction}

\noindent The production, application, and/or measurement of polarised X-/gamma rays are of significant interest to the fields of synchrotron science \cite{Paganin2006,Panaccione2009,Strocov2010,Yamamoto2014} and X-/gamma-ray astronomy \cite{Novik1975,Meszaros1988,Lei1997,Kumar2015,McConnell2017}. In synchrotron science specialised equipment and end-stations are required to exploit the properties of polarised X-/gamma ray interactions to modify the state and/or probe the structure of a target object \cite{Strocov2014,Kummer2016,Hoesch2017}. In the case of X-/gamma-ray astronomy, coded aperture, Compton and pair production telescopes are utilised to determine specific structural properties of astronomical bodies through the measurement of the relative polarisation strength of the their emitted X-/gamma ray radiation \cite{Schonfelder1993,Lin2003,Ubertini2003,Vedrenne2003,Kamae2008,Singh2014,Giomi2017}. In both of these fields the design, development and optimsation processes of their respective equipment typically relies on the use of Monte Carlo radiation transport modelling toolkits \cite{Zoglauer2006,Agapov2009,Bulgarelli2012,Chattopadhyay2013,Cornelius2014,Dimmock2015,Moiseev2015,Chauvin2016}. Of the available Monte Carlo radiation transport modelling toolkits Geant4 \cite{G42003,G42006,G42016} is the most commonly utilised for these tasks due to its flexible nature and wide array of polarised radiation physics transport models \cite{G4Phys2020}.

The Geant4 toolkit for the simulation of the passage of particles through matter is the result of a world wide collaboration of over 100 scientists and software engineers that has spanned the last 26 years \cite{G41994}. To date a total of ten versions have been released, with each release incrementally improving and optimising the core tracking, geometry and hits collection architecture. At the same time new particle types and physics models, including electromagnetic, hadronic and optical processes, spanning energies of a few eV to hundreds of TeV have been added to increase the functionality of Geant4 \cite{G42003,G42006,G42016,G4Phys2020}. In the release of Geant4 version 8.2 an extensive set of polarised particle transport models were added to complement the existing models for polarised X-/gamma ray photoelectric absorption, Compton scattering, and Rayleigh scattering \cite{Schalick2007}. Since then new polarised X-/gamma rays transport models have been added for gamma conversion/pair production in 2009 \cite{Depaola2009} and 2018 \cite{Bernard2018,Semeniouk2019}, Compton scattering in 2016 \cite{G4Phys2020}, and elastic scattering (Rayleigh, Delbr\"{u}ck and Nuclear Thomson scattering) in 2019 \cite{Omer2019}.

The present work outlines the reconfiguration of the ``G4LowEPPhysics" electromagnetic physics constructor for polarised X-/gamma ray transport in Geant4. This electromagnetic physics constructor from Geant4 version 11.0 aims to offer users an easy way to implement the ``best set" of electromagnetic physics models for studies exploring the transport of low energy polarised X-/gamma rays. Section \ref{sec:M} contains an overview of the physics models implemented in ``G4LowEPPhysics" and outlines it's experimental validation against Compton X-ray polarimetry measurements from the BL38B1 beamline at the SPring-8 synchrotron (Sayo, Japan). The results and discussion from this experimental validation, and an overall conclusion then follow in Sections \ref{sec:RD} and \ref{sec:C} respectively.

\section{Method}
\label{sec:M}

\subsection{Overview of the ``G4LowEPPhysics" Electromagnetic Physics Constructor}
The electromagnetic physics constructor ``G4LowEPPhysics" aims to offer the ``best set" of electromagnetic physics models available in Geant4 for low energy polarised X-/gamma ray transport. It was developed from the library of available electromagnetic physics ``Model Classes" of Geant4 version 11.0 and will be reviewed with each new Geant4 release. A list of the X-/gamma ray, electron and positron physics ``Model Classes" implemented in ``G4LowEPPhysics" for each physical process with their respective energy activation range in Geant4 version 11.0 is displayed in Table \ref{tab:tab1}. Whilst it can be seen that the applicable energy range of ``G4LowEPPhysics" spans the ev to TeV scale, these ``Model Classes" were selected as they are known to be based on theoretical models that possess high physical accuracy below 10 MeV \cite{G42016,G4Phys2020}. In addition these models are also applicable to the simulation of non-polarised X-/gamma rays as during their first interaction a random axis of polarisation is sampled if not defined at creation/emission, and then tracked for remainder of the X-/gamma ray's propagation. Finally, with ``G4LowEPPhysics" both X-ray fluorescence and Auger electron emission from atomic deexcitation are enabled by default. More information on these ``Model Classes'' can be found in the references outlined in Table 1 and the Geant4 Physics Reference Manual \cite{G4Phys2020}.

\begin{table}[tbh]
\centering
\begin{tabular}{|c|c|c|c|}

\hline
Physical & Model & Energy & References \\
Process  & Class & Range    & \\
\hline
\hline
\multicolumn{4}{|c|}{\textbf{X-/Gamma ray}} \\
\hline
Photoelectric Absorption &  G4LivermorePolarizedPhotoElectricModel  & eV - TeV & \cite{Gavrila1959,Gavrila1961}  \\

Rayleigh Scattering & G4LivermorePolarizedRayleighModel & 250 eV - TeV & \cite{Depaola2003}\\
Compton Scattering & G4LowEPPolarizedComptonModel & eV - TeV & \cite{Brown2014} \\
Gamma Conversion & G4BetheHeitler5DModel  & 1.022 MeV - TeV & \cite{Bernard2018,Semeniouk2019} \\
\hline
\multicolumn{4}{|c|}{\textbf{Electron}} \\
\hline

\multirow{2}{*}{Ionisation} & G4LivermoreIonisationModel & eV - 100 keV   & \cite{Seltzer1989}  \\
                          & G4UniversalFluctuation     & 100 keV - TeV  & \cite{LassilaPerin1995} \\
\multirow{2}{*}{Bremsstrahlung}  & G4SeltzerBergerModel & eV - GeV & \cite{Allison2012} \\
                & G4eBremsstrahlungRelModel & GeV - TeV & \cite{Allison2012} \\
\multirow{2}{*}{Multiple Scattering} & G4GoudsmitSaundersonMscModel & eV - 100 MeV & \cite{Bagulya2017} \\
 & G4WentzelVIModel & 100 MeV - TeV & \cite{Ivanchenko2010} \\

\hline
\multicolumn{4}{|c|}{\textbf{Positron}} \\
\hline

Annihilation & G4eplusAnnihilation & 1.022 MeV - TeV & \cite{Heitler1954,Nelson1985} \\
\multirow{2}{*}{Ionisation}  & G4PenelopeIonisationModel & eV - 100 keV & \cite{Salvat2008} \\
                          & G4UniversalFluctuation     & 100 keV - TeV  & \cite{LassilaPerin1995} \\
\multirow{2}{*}{Bremsstrahlung}  & G4SeltzerBergerModel & eV - GeV & \cite{Allison2012} \\
                & G4eBremsstrahlungRelModel & GeV - TeV & \cite{Allison2012} \\
\multirow{2}{*}{Multiple Scattering} & G4GoudsmitSaundersonMscModel & eV - 100 MeV & \cite{Bagulya2017} \\
 & G4WentzelVIModel & 100 MeV - TeV & \cite{Ivanchenko2010} \\
 
\hline
\end{tabular}
 \caption[]{The X-/gamma ray, electron and positron physics models classes and their respective energy activation range implemented for each physical process in ``G4LowEPPhysics". Detail descriptions of these ``Model Classes'' can be found in their respective references above and the Geant4 Physics Reference Manual \cite{G4Phys2020}.}
\label{tab:tab1}
\end{table}

\subsection{Experimental Validation of ``G4LowEPPhysics"}

Experimental validation of the ``G4LowEPPhysics" electromagnetic physics constructor was undertaken using the Compton X-ray polarimetry measurement data collected at the SPring-8 synchrotron BL38B1 beamline reported in Tokanai et al. \cite{Tokanai2004}. The following provides an overview of the experimental setup and measurements undertaken by Tokanai et al. \cite{Tokanai2004}, outlines the development of a Geant4 application based on the geometry of the SPring-8 synchrotron BL38B1 beamline, and describes the simulation parameters that were implemented to experimentally validate ``G4LowEPPhysics".

\subsubsection{SPring-8 Synchrotron BL38B1 Beamline Experimental Setup and Measurements}
\label{sec:M:Exp}

Tokanai et al. constructed an X-ray polarimeter from two cadmium telluride (CdTe) detectors (Amptek XR-100T-CdTe) and a custom built vertically orientated acrylic rotational stage \cite{Tokanai2004}. These two CdTe detectors were mounted onto the rotational stage perpendicular to one another, and focused towards a scattering sample stage that is located at the centre of rotation. Each detector was placed on an independent linear travel stage that enabled the radial distance between their detection window and the scattering target to be controlled in order to optimise the scattered radiation detection sensitivity. The whole unit was placed onto the BL38B1 beamline with the central beam path orientated perpendicular to the plane defined by these three elements, and focused at the centre of the scattering sample stage \cite{Tokanai2004}.

The orientation of the three key elements of the X-ray polarimeter, the scattering target and both CdTe detectors, with respect to other detector and filtering elements utilised on the BL38B1 beamline can be seen in Figure \ref{fig:1}. In this experimental configuration a near 100\% horizontally polarised 100 {\textmu}m by 100 {\textmu}m collimated mono-energetic X-ray beam (represented via the dashed line) propagated left to right along the z-axis from the beam defining slits to the surface of the scattering target through a He gas flight tube. Relative to the incident beam, the two CdTe detectors were orientated at a 90 degree radial scattering angle ($\theta = 90 ^{\circ}$) 150 mm away from the centre of the sample scattering stage \cite{Tokanai2004}. Through the vertically orientated acrylic rotational stage these two CdTe detectors were able to measure the azimuthal angular scattering ($\phi$) modulation of the BL38B1 beamline's mono-energetic X-ray beam due to its horizontal plane polarisation. In addition to the X-ray polarimeter, Figure \ref{fig:1} illustrates that a 170 mm long He gas-flow type ionisation chamber, set of metal filters, and custom YAP(Ce) scintillator based energy discriminating photon counting detector were present downstream 700 mm, 920 mm, and 1170 mm from the centre of the scattering sample stage respectively (Tokanai~F. \textit{Personal Communication}, March 2020). Whilst the He gas-flow type ionisation chamber was present in the experimental configuration, none of its measured data was utilised or reported in Tokanai et al. \cite{Tokanai2004}.

A total of three different sets of experimental measurements were undertaken and reported in Tokanai et al. \cite{Tokanai2004} to observe the azimithal scattering modulation of the linearly polarised X-ray BL38B1 beamline. Two of these experimental measurement sets were collected for a 20 keV mono-energetic X-ray beam using two different cylindrical polypropylene scattering targets placed with their circular face perpendicular to the incident beam. The first cylindrical polypropylene scattering target (S1) had a 10 mm diameter and 10 mm length, whereas the second (S2) had a 5 mm diameter and 7 mm length \cite{Tokanai2004}. In both 20 keV mono-energetic X-ray beam experimental measurements a 300 {\textmu}m thick copper filter was used to attenuate the X-ray beam before striking the YAP(Ce) scintillator photon counting detector to avoid detector saturation and limit dead-time. The energy window of the YAP(Ce) scintillator photon counting detector was optimised to target the 20 keV X-ray photopeak signal, and the two CdTe detectors obtained the azimithal angle ($\phi$) modulated X-ray scattering spectra at 15 degree intervals over a 360 degree range \cite{Tokanai2004}. The third, and final, set experimental measurements was undertaken for a 40 keV mono-energetic X-ray beam using the second cylindrical polypropylene scattering target S2. In this set of experimental measurements a 600 {\textmu}m thick lead filter was used to attenuate the X-ray beam before striking the YAP(Ce) scintillator photon counting detector (Tokanai~F. \textit{Personal Communication}, March 2020). As with the two previous experimental measurements, the YAP(Ce) scintillator photon counting detector energy window was optimsed to target the 40 keV X-ray photopeak signal and two CdTe detectors obtained the azimithal angle ($\phi$) modulated X-ray scattering spectra in 15 degree intervals over a 360 degrees range \cite{Tokanai2004}. 

\subsubsection{SPring-8 Synchrotron BL38B1 Beamline Geant4 Application}
\label{sec:M:G4D}
\begin{figure}[tbh!]   
   \centering
   \includegraphics[width=1.0\textwidth]{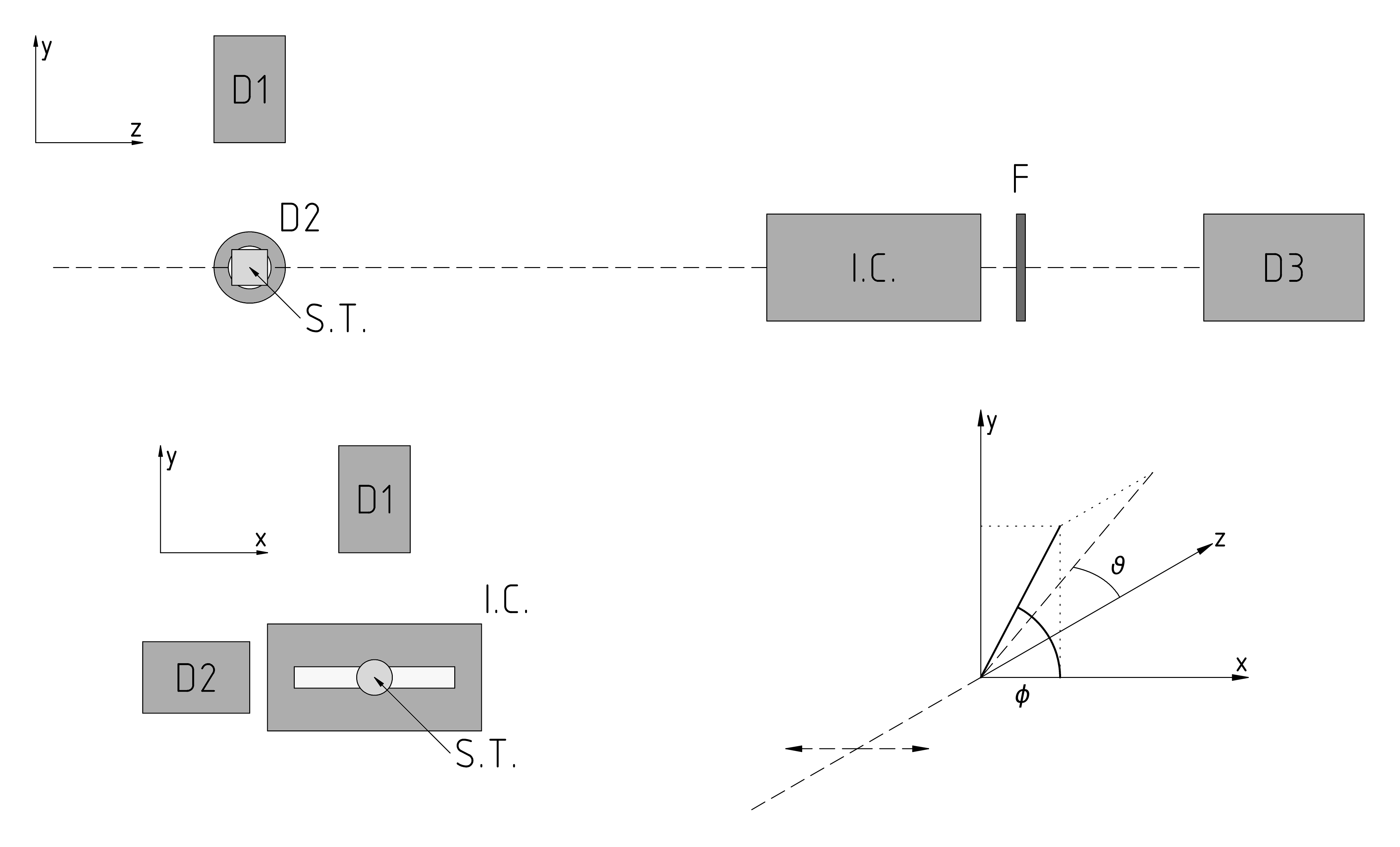}
\caption{Orientation of the three key elements of the X-ray polarimeter, the scattering target (S.T.) and both CdTe detectors (D1 and D2), with respect to the 170 mm long He gas-flow type ionisation chamber (I.C.), set of metal filters (F), and YAP(Ce) scintillator based energy discriminating photon counting detector (D3) locations at the BL38B1 beamline (SPring-8) in the y-z (top) and x-y (bottom left) planes \cite{Tokanai2004}. The dashed line represents the propagation of the near 100\% linear horizontally polarised 100 {\textmu}m by 100 {\textmu}m collimated mono-energetic X-ray beam along the z-axis, and the scattering coordinate system of the experiment setup is displayed in the bottom right panel. Tokanai et al. measured the azimuthal angular modulation of the scattered polarised X-ray beam with D1 and D2 150 mm away from the centre of the S.T. for $\theta = 90 ^{\circ}$ and over a range of $0^{\circ}\leq\phi\leq 360^{\circ}$ in steps of 15$^{\circ}$.}
\label{fig:1}
\end{figure}

Geant4 version 10.6 was used to construct an application of the experimental SPring-8 synchrotron BL38B1 beamline setup illustrated in Figure \ref{fig:1}. In this Geant4 application a 100\% horizontally polarised 100 by 100 {\textmu}m collimated mono-energetic X-ray source was implemented originating 10 cm from the centre of the scattering sample stage in a ``experimental hall" filled with air. Six different objects were simulated in the ``experimental hall" to mimic the geometry seen in Figure \ref{fig:1} that can be classified as one of two general object types: ``structural objects" and ``active radiation detectors". ``Structural objects" are objects that were present along the collimated mono-energetic X-ray beam path (represented via the dashed line in Figure \ref{fig:1}) and that did not yield any experimental data that was utilised/reported in Tokanai et al. \cite{Tokanai2004}. This includes the scattering target, He gas-flow type ionisation chamber, and the set of metal filters. For the scattering target a cylindrical volume composed of solid polypropylene was implemented with the ability to modify its diameter and length to match that of S1 and S2. In the case of the He gas-flow type ionisation chamber a 5 mm thick stainless steel box of dimensions 110 $\times$ 50 $\times$ 170 mm, with 50 {\textmu}m thick Kapton film 80 mm $\times$ 10 mm entrance/exit windows, filled with He gas at standard laboratory condition was implement to mimic the unit available at the BL38B1 beamline (Tokanai~F. \textit{Personal Communication}, March 2020). Whereas for the set of metal filter a single slab of solid variable material composition, i.e. copper or lead, and thickness with a cross-sectional surface area of 50 mm $\times$ 50 mm was implemented orientated perpendicular to the X-ray beam.

Three ``active radiation detectors" were implemented: two Amptek XR-100T-CdTe detectors, and a custom YAP(Ce) scintillator based energy discriminating photon counting detector. For the Amptek XR-100T-CdTe detectors the geometry and material composition of the CdTe detection element, peltier cooler, mounting stud, Beryllium window, and Ni metal housing were implemented based on the 3 mm $\times$ 3 mm $\times$ 1 mm  CdTe chip version outlined in the online Amptek XR-100T-CdTe documentation (\href{https://www.amptek.com/products/x-ray-detectors/cdte-x-ray-and-gamma-ray-detectors/xr-100cdte-x-ray-and-gamma-ray-detector}{https://www.amptek.com}). Modelling of the Amptek XR-100T-CdTe detector's energy response and output spectra was implemented in a three step process under and it was the assumed that there was no pulse pile up due to multiple primary X-ray detection or electronics dead-time. In the first step the charge collection efficiency of each energy deposition location for a given incident X-ray within the CdTe chip was modelled relative to its electrode structure with the Hecht relation:
\begin{equation}
    \displaystyle \eta(x) = \left[\frac{\lambda_{e}}{T}\left(1 - e^{-\frac{x}{\lambda_{e}}}\right)  + \frac{\lambda_{h}}{T}\left(1 - e^{-\frac{T-x}{\lambda_{h}}}\right)   \right]
\end{equation}

\noindent where $T=1$ mm is the thickness, $\lambda_{e}=132$ mm is the electron trapping length, $\lambda_{h}=8$ mm is the hole trapping length, and $x$ is the relative interaction depth of the incident X-ray within the CdTe chip active volume \cite{Redus2007}. For the second step, the total energy deposited within the CdTe chip's active region was then calculated for each simulated primary X-ray through the weighted summation of the charge collection efficiency of each energy deposition location:

\begin{equation}
    \displaystyle E_{n} = \sum_{i=1}^{n}\eta_{i(x)}E_{i} 
\end{equation}

\noindent where $E_{i}$ is the energy deposited at each interaction location within the CdTe chip's active region. The total energy deposited within the CdTe chip's active region was then blurred using a Gaussian function of 0.45 keV full width at half maximum (FWHM). 

In the case of the custom YAP(Ce) scintillator based energy discriminating photon counting detector, it was constructed through taping a solid puck of YAP(Ce) onto the surface of a 3 inch diameter Photomultiplier Tube (PMT) with 3M Scotch Super 88 Premium Vinyl Electrical Tape and read out using a single channel analyser (SCA) with an approximate low energy window threshold of 10 keV (Tokanai~F. \textit{Personal Communication}, March 2020). To approximate it's geometry and material composition, a butted 50 mm diameter-1 mm thick puck of solid YAP(Ce) and 76 mm diameter-1 mm thick glass puck was implemented inside of a 76.88 mm diameter-2.88 mm thick solid Polyvinylchloride (PVC) cylinder. The SCA based electronics readout was modelled through the use of a simple logic counter with a 10 keV low energy window threshold. Again, it was assumed that there was no pulse pile up due to multiple primary X-ray detection or electronics dead-time.

\subsection{Experimental Validation Simulation Parameters}
\label{sec:M:G4S}

\begin{table}[tbh]
\centering
\begin{tabular}{|c|c|c|c|c|c|}

\hline
Simulation & Incident X-ray & Polypropylene         & Metal Filter & Metal Filter & Detector Azimithal \\
Set        & Energy         & Scattering Target & Material     & Thicknesses   & Angle Positions \\
\hline
\hline
\multirow{2}{*}{1}         & \multirow{2}{*}{20 keV}         & 10 mm diameter, & \multirow{2}{*}{Copper (Cu)} & 300 \textmu m & \multirow{2}{*}{$0^{\circ}, 15^{\circ}, 30^{\circ}, ... , 180^{\circ}$} \\
          &        & 10 mm length (S1) &  & $\pm$ 25, 50 \textmu m & \\
\hline
\multirow{2}{*}{2}         & \multirow{2}{*}{20 keV}         & 5 mm diameter, & \multirow{2}{*}{Copper (Cu)} & 300 \textmu m & \multirow{2}{*}{$0^{\circ}, 15^{\circ}, 30^{\circ}, ... , 180^{\circ}$} \\
          &        & 7 mm length (S2) &  & $\pm$ 25, 50 \textmu m & \\
\hline
\multirow{2}{*}{3}         & \multirow{2}{*}{40 keV}         & 5 mm diameter, & \multirow{2}{*}{Lead (Pb)} & 600 \textmu m & \multirow{2}{*}{$0^{\circ}, 15^{\circ}, 30^{\circ}, ... , 180^{\circ}$} \\
          &        & 7 mm length (S2) &  & $\pm$ 25, 50 \textmu m & \\          
\hline
\end{tabular}
 \caption[]{Simulation parameters for the three different sets of simulations that were undertaken with the developed Geant4 BL38B1 beamline model and ``G4LowEPPhysics" electromagnetic physics constructor. Here the set of detector azimithal angle positions was limited to range of $0^{\circ}\leq\phi\leq 180^{\circ}$ in steps of 15$^{\circ}$ to exploit the scattering symmetry around the plane of X-ray source polarisation (i.e. the horizontal plane) and reduce the extent of required computational resources.}
\label{tab:tab2}
\end{table}

Table \ref{tab:tab2} presents a summary of the parameters for the three different sets of simulations that were undertaken with the developed Geant4 BL38B1 beamline model and ``G4LowEPPhysics" electromagnetic physics constructor. In each simulation set the incident mono-energetic X-ray energy, scattering target, and metal filters were modified to match the experimental configurations of one of the three different experimental measurements reported in Tokanai et al. \cite{Tokanai2004} (see Section \ref{sec:M:Exp}). For each of these different incident mono-energetic X-ray energy, scattering target, and metal filter configurations two geometrical parameters were varied: 1) the relative Amptek XR-100T-CdTe detector's azimithal angular ($\phi$) orientations over a range of $0^{\circ}\leq\phi\leq 180^{\circ}$ in steps of 15$^{\circ}$ (see Figure \ref{fig:1}); and 2) the metal filter thicknesses over a range of $\pm$ 50 \textmu m in steps of 25 \textmu m around their stated thicknesses in Section \ref{sec:M:Exp}. These two parameter sweeps were undertaken to: 1) emulate the procedure employed to measure the reported experimental azimithal angle ($\phi$) modulated X-ray scattering spectra and normalised CdTe to YAP(Ce) detector signal ratios in Tokanai et al. \cite{Tokanai2004}\footnote{Note that the simulated normalised CdTe to YAP(Ce) detector signal ratios were calculated in the same manner as the experimental data outlined in Tokanai et al. \cite{Tokanai2004}: i.e. the sum of the CdTe spectra channels above 10 keV divided by the logic counter output of the YAP(Ce) detector.}; and 2) illustrate the accuracy of ``G4LowEPPhysics" validation with respect to one of the more significant sources of experimental uncertainty (effective metal filter thickness due to alignment accuracy and manufacturing tolerances). Finally, for each CdTe detector azimithal angle ($\phi$) and metal foil thickness configuration 10$^{10}$ primary X-rays were simulated using a maximum particle step length of 10 {\textmu}m and a low-energy particle cut off of 250~eV.

\section{Results and Discussion}
\label{sec:RD}

\begin{figure}[tbh]    
    \centering
    \begin{subfigure}
    \centering 
        \includegraphics[width=0.5\textwidth, trim = {5 0 30 20},clip]{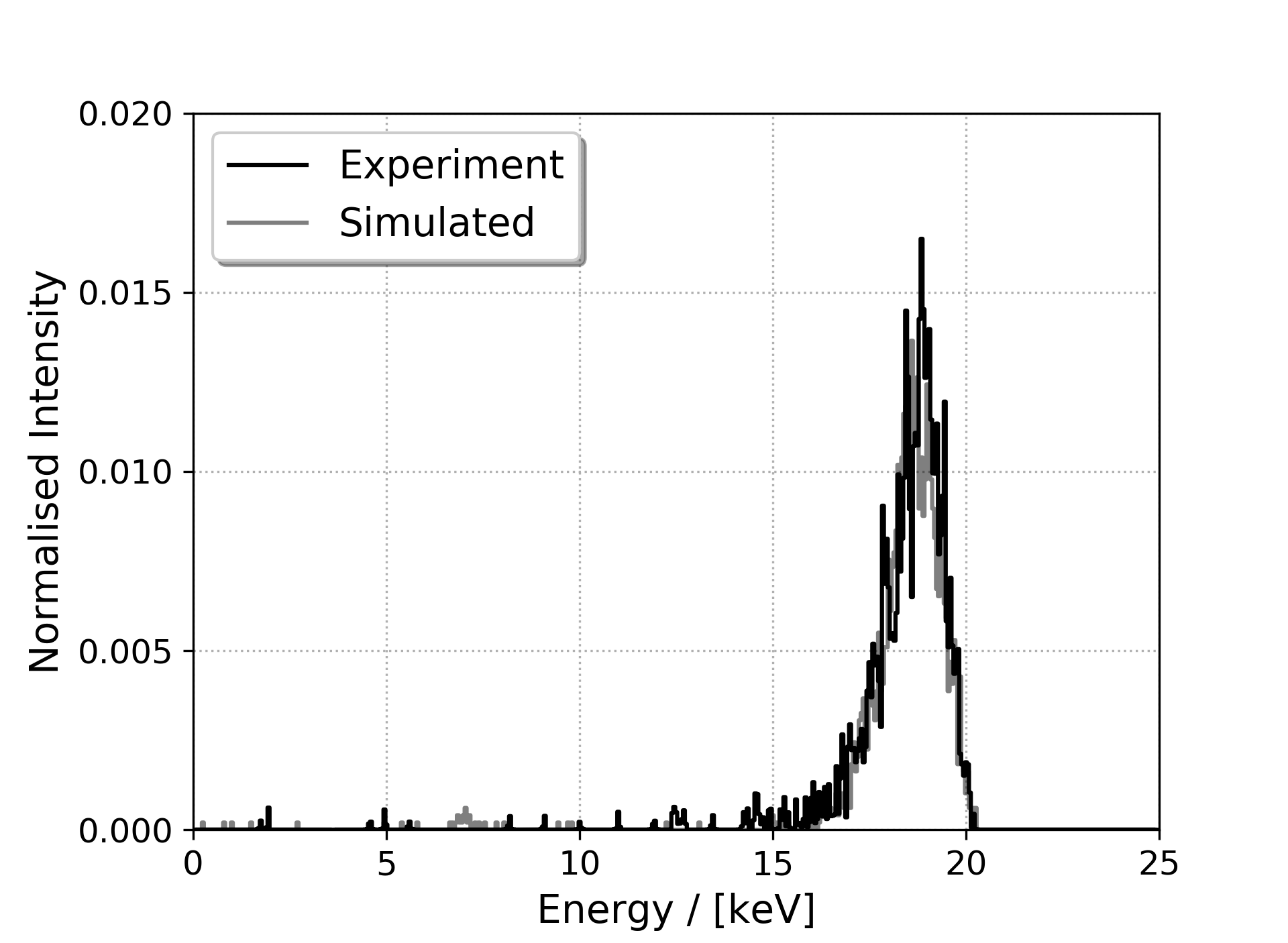}
        \label{fig:r1a}
    \end{subfigure}
    \begin{subfigure}
    \centering
        \includegraphics[width=0.5\textwidth, trim = {5 0 30 20},clip]{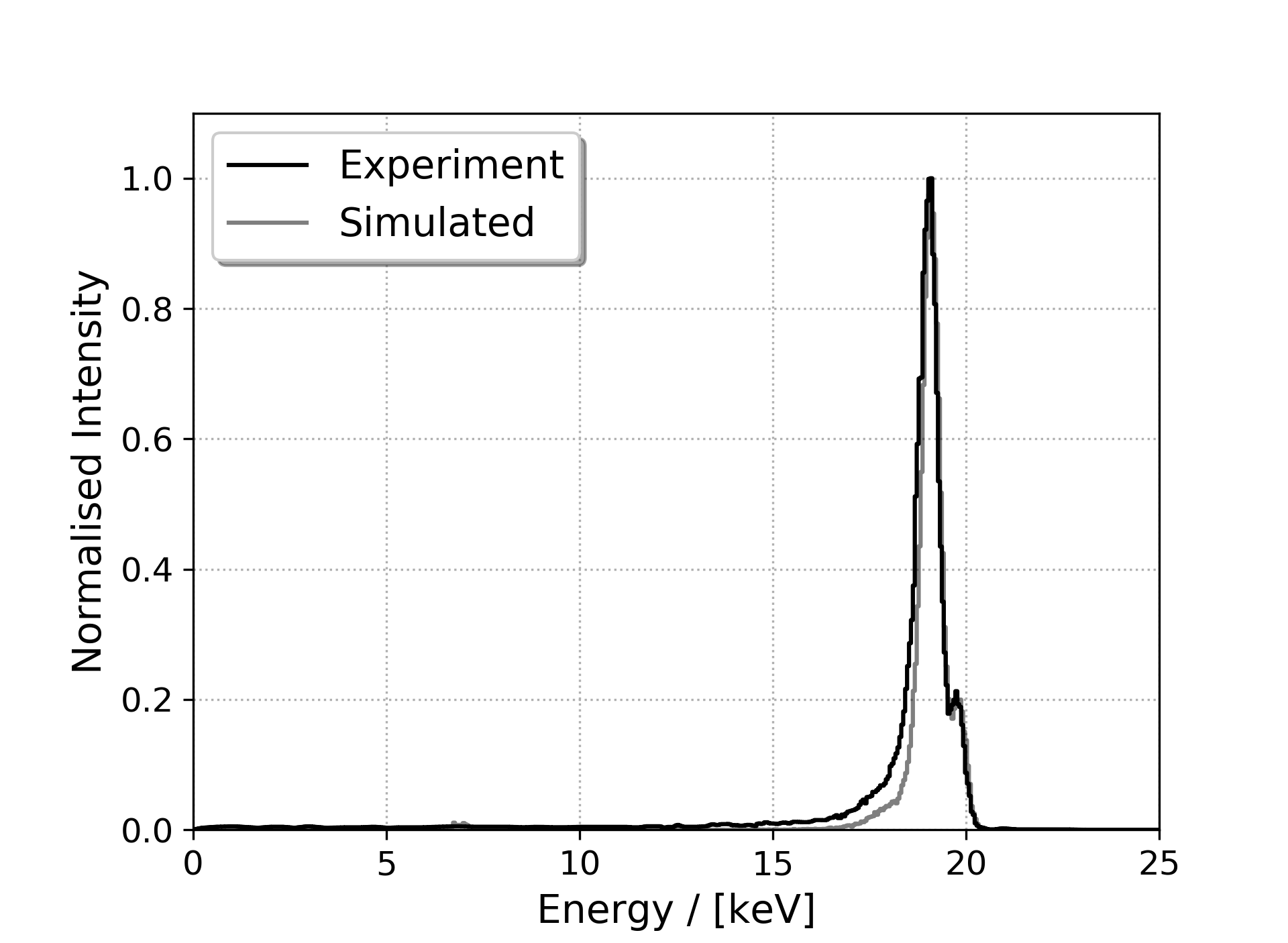}
        \label{fig:r1b}
    \end{subfigure}

    \begin{subfigure}
    \centering 
        \includegraphics[width=0.5\textwidth, trim = {5 0 30 20},clip]{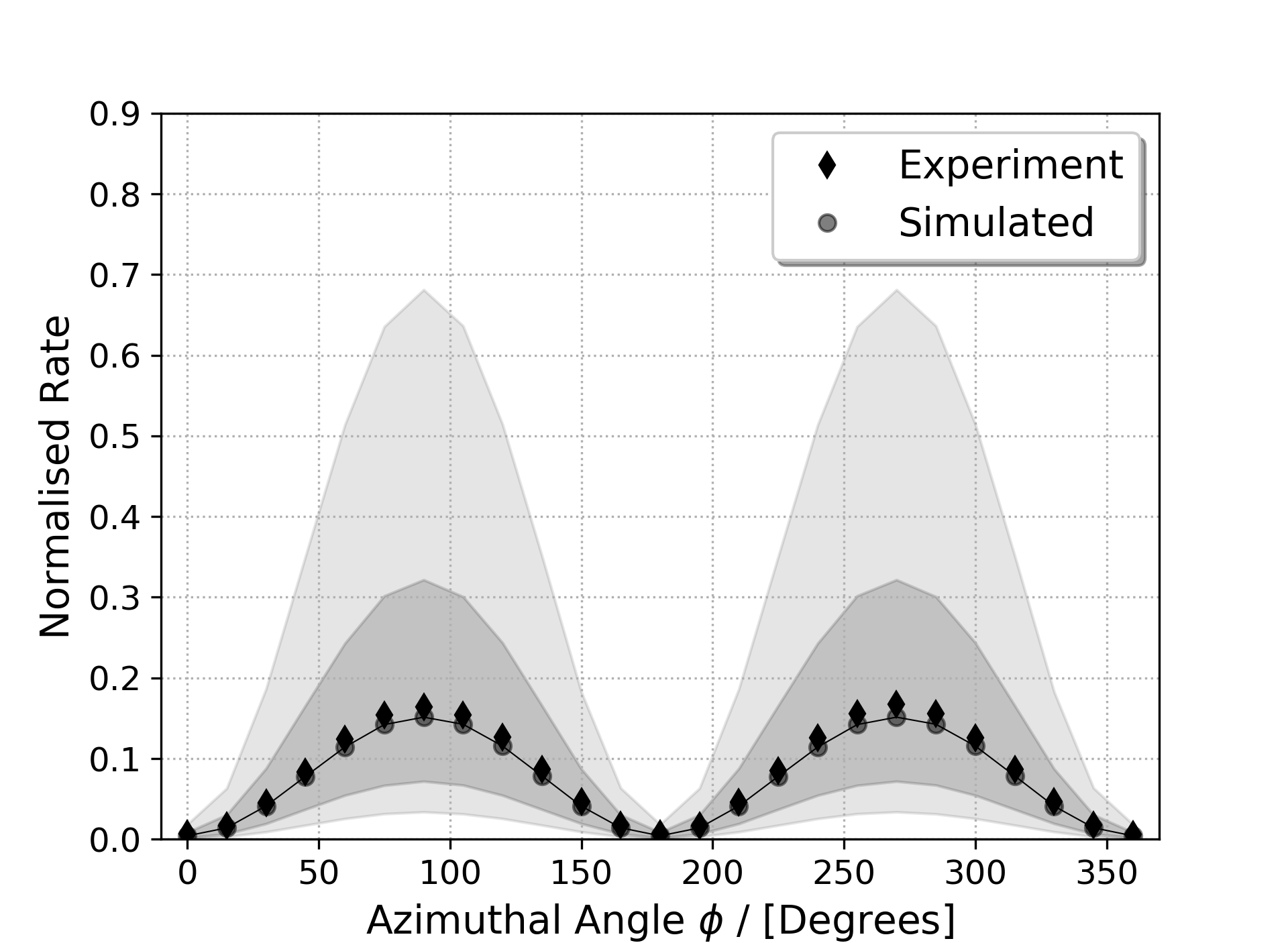}
        \label{fig:r1c}
    \end{subfigure}
\caption{Comparison of the experimental and simulation output for Simulation Set 1 (20 keV incident X-ray energy, scattering target S1, and a Cu metal filter) in Table \ref{tab:tab2}. Three different data-sets are displayed: the Amptek XR-100T-CdTe detector scattered X-ray energy spectra at $\phi=0^{\circ}$ (top), the Amptek XR-100T-CdTe detector scattered X-ray energy spectra at $\phi=90^{\circ}$ (middle), and the azimithal angle ($\phi$) normalised CdTe to YAP(Ce) detector signal ratios (bottom). In (bottom) the markers and solid black line represent the ideal filter thickness simulated data, with the two grey inner and outer bands corresponding to the $\pm$ 25 \textmu m and $\pm$ 50 \textmu m filter simulation outputs respectively.}
\label{fig:r1}
\end{figure}

The results from the experimental and developed Geant4 application simulation output with the ``G4LowEPPhysics" electromagnetic physics constructor for Simulation Set 1 in Table \ref{tab:tab2} (20 keV incident X-ray energy, scattering target S1, and a Cu metal filter) is presented in Figure \ref{fig:r1}. Three different data-sets are displayed in Fig. \ref{fig:r1}: the Amptek XR-100T-CdTe detector scattered X-ray energy spectra at $\phi=0^{\circ}$ (top), the Amptek XR-100T-CdTe detector scattered X-ray energy spectra at $\phi=90^{\circ}$ (middle), and the azimithal angle ($\phi$) normalised CdTe to YAP(Ce) detector signal ratios for $0^{\circ}\leq\phi\leq 360^{\circ}$ in steps of 15$^{\circ}$ (bottom). Here, both the experimental and simulated energy spectra sets are normalised with respect to their maximum intensity of the $\theta\approx90^{\circ}$ Compton scattered X-ray peak seen in Fig. \ref{fig:r1} (middle). Comparison of the experimental and simulation energy spectra for $\phi=0^{\circ}$ (Fig. \ref{fig:r1} (top)) illustrate a high level of correlation between the combined $\theta\approx90^{\circ}$ Rayleigh scattering and Compton scattered X-ray peaks. In these energy spectra the level of correlation is such that even the statistical variance within the energy spectra is near identical. When a similar comparison is undertaken for Fig. \ref{fig:r1} (middle) ($\phi=90^{\circ}$), a similarly high level of correlation between the two energy spectra can be observed with the exception that the Compton scattered X-ray peak tail is reduced in the simulated spectra. This observed reduction in Compton scattered X-ray peak tail in the simulated spectra can be attributed to the first order approach that was employed in modelling the charge collection process in the CdTe chip \cite{Redus2007}. However even with the use of the first order approach, the normalised intensity of all features of the energy spectra obtained with the ``G4LowEPPhysics" electromagnetic physics constructor are a near exact match to the experimental energy spectra presented in Fig. \ref{fig:r1} (top) and (middle).

Figure \ref{fig:r1} (bottom) contains the experimental and simulated azimithal angle ($\phi$) normalised CdTe to YAP(Ce) detector signal ratios for $0^{\circ}\leq\phi\leq 360^{\circ}$ in steps of 15$^{\circ}$. In the case of the simulated data, five different thickness of the metal filter were explored with the value matching that stated for the experimental measurement of Tokanai et al. \cite{Tokanai2004} (Cu: 300 \textmu m) represented via the solid circle markers. The other four explored metal filter thickness are separated into two groups, 300 $\pm$ 25 \textmu m (275 \textmu m and 325 \textmu m) and 300 $\pm$ 50 \textmu m (250 \textmu m and 350 \textmu m), to emulate the potential variations in the effective metal filter thickness due to experimental alignment accuracy and manufacturing tolerances. In Fig. \ref{fig:r1} (bottom) the $\pm$ 25 \textmu m and $\pm$ 50 \textmu m simulation results are indicated by the dark-shaded and light-shaded region bounds respectively. Inspection of Fig. \ref{fig:r1} (bottom) illustrates that a high level of correlation exists between the experimental (solid diamonds) and the 300 \textmu m thick Cu metal filter thickness simulation (solid circle) normalised CdTe to YAP(Ce) detector signal ratios. It can be seen that across the entire azimithal angle ($\phi$) range the experimental diamond markers overlap with their respective 300 \textmu m simulation circle markers, and, in turn, fall well within the $\pm$ 25 \textmu m shaded region that represents a less than $\pm$ 10 \% difference from the true simulated metal filter thickness.  

\begin{figure}[tbh]    
    \centering
    \begin{subfigure}
    \centering 
        \includegraphics[width=0.6\textwidth, trim = {5 0 30 20},clip]{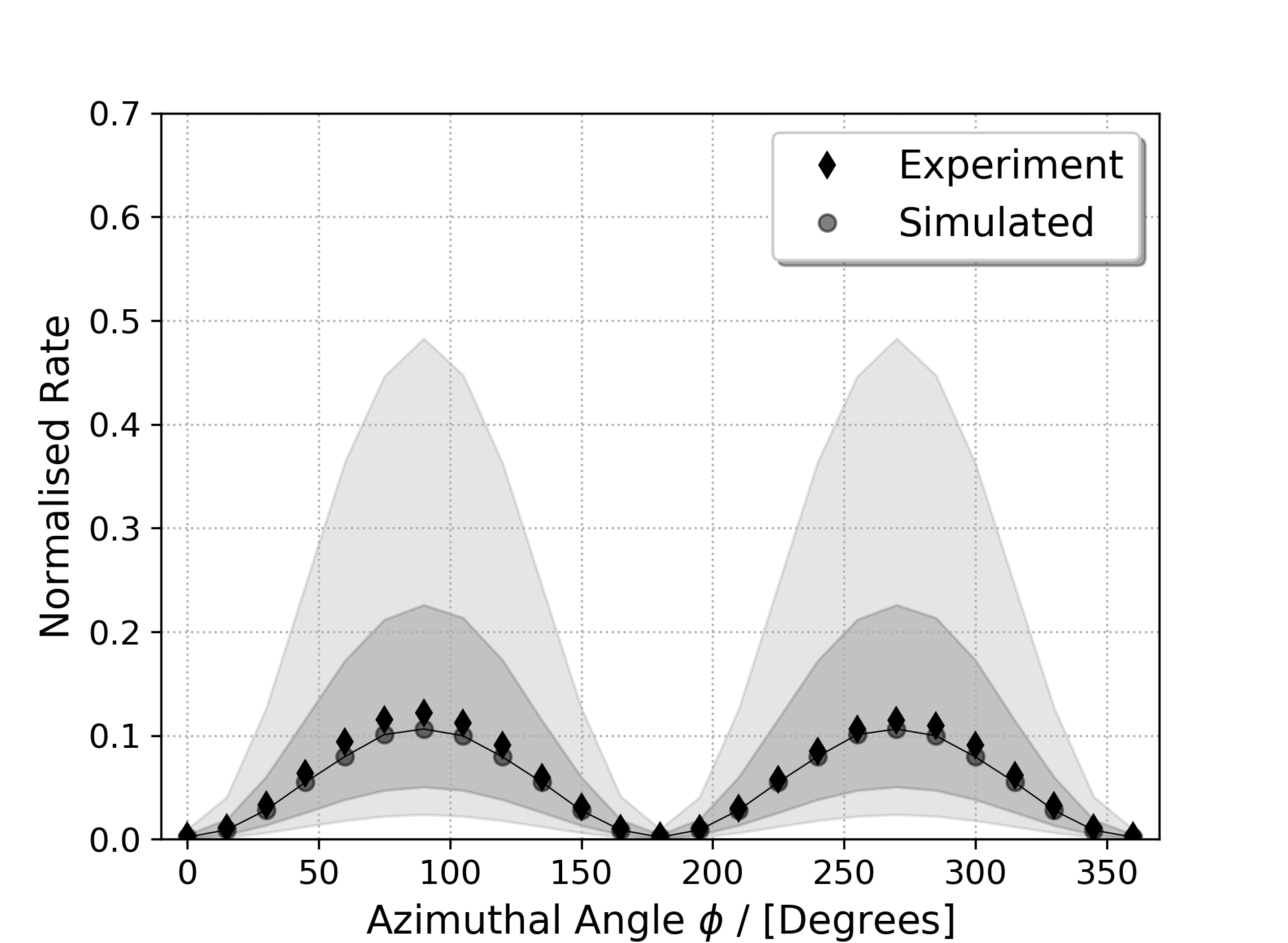}
        \label{fig:r2a}
    \end{subfigure}
    \begin{subfigure}
    \centering
        \includegraphics[width=0.6\textwidth, trim = {5 0 30 20},clip]{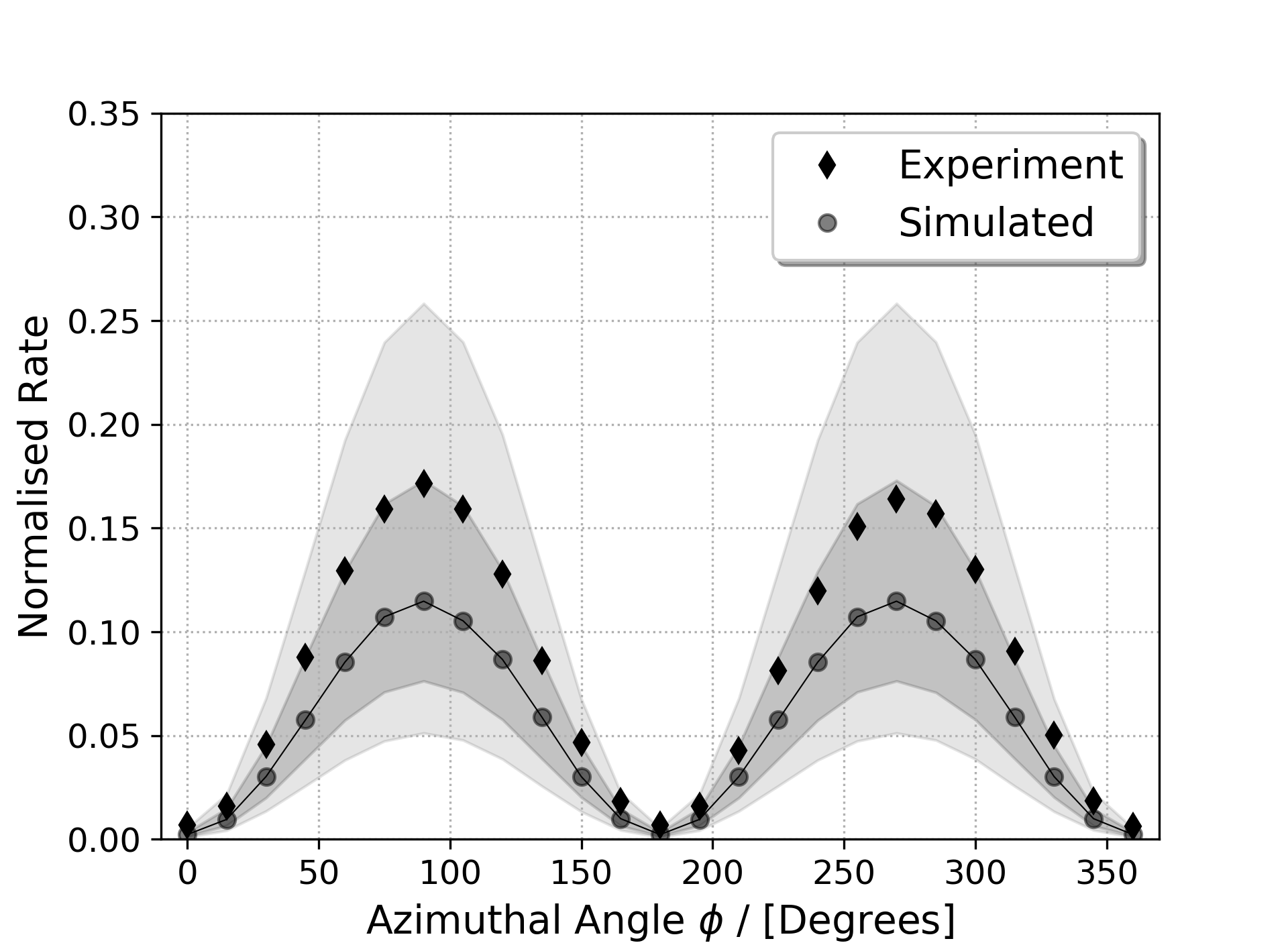}
        \label{fig:r2b}
    \end{subfigure}
\caption{Comparison of the experimental and simulation azimithal angle ($\phi$) normalised CdTe to YAP(Ce) detector signal ratios for Simulation Set 2 and 3 in Table \ref{tab:tab2}. Here the (top) panel contains the experimental and simulation data for Simulation Set 2 (20 keV incident X-ray energy, scattering target S2, and a Cu metal filter), and the (bottom) panel contains the experimental and simulation data for Simulation Set 3 (40 keV incident X-ray energy, scattering target S2, and a Pb metal filter). The markers and solid black line represent the ideal filter thickness simulated data, with the two grey inner and outer bands corresponding to the $\pm$ 25 \textmu m and $\pm$ 50 \textmu m filter simulation outputs respectively.}
\label{fig:r2}
\end{figure}

The experimental and simulated azimithal angle ($\phi$) normalised CdTe to YAP(Ce) detector signal ratios for Simulation Set 2 and 3 outlined in Table \ref{tab:tab2} is presented in Fig. \ref{fig:r2}. Figure \ref{fig:r2} (top) contains the experimental and simulation data for Simulation Set 2 (20 keV incident X-ray energy, scattering target S2, and a Cu metal filter), and Fig. \ref{fig:r2} (bottom) contains the experimental and simulation data for Simulation Set 3 (40 keV incident X-ray energy, scattering target S2, and a Pb metal filter). In both sets of simulated result the metal thickness values stated in Tokanai et al. \cite{Tokanai2004} (Cu: 300 \textmu m and Pb: 600 \textmu m) are represented via the solid circle markers, and the $\pm$ 25 \textmu m and $\pm$ 50 \textmu m simulation results through the dark-shaded and light-shaded region bounds respectively. Inspection of the Simulation Set 2 results in Fig. \ref{fig:r2} (top) illustrates that a high level of correlation exists between the experimental (solid diamonds) and simulation (solid circle) normalised CdTe to YAP(Ce) detector signal ratios. As with the Simulation Set 1 presented in Fig. \ref{fig:r1} (bottom), it can be seen that across the entire $\phi$ range the experimental diamond markers overlap with their respective 300 \textmu m simulation circle markers, and, in turn, fall well within the $\pm$ 25 \textmu m shaded region that represents a less than $\pm$ 10 \% difference from the true simulated metal filter thickness. Whereas for the the Simulation Set 3 presented in  Fig. \ref{fig:r2} (bottom), the level of the correlation between the experimental (solid diamonds) and 600 \textmu m thick Pb filter simulation (solid circle) is weaker with the experimental results aligning strongly with the upper bound of the $\pm$ 25 \textmu m shaded region. However, it should be noted that in the case of the Simulation Set 3 results the $\pm$ 25 \textmu m shaded region represents a less than $\pm$ 5 \% difference from the true simulated metal filter thickness.  

The results presented above illustrate a high level of correlation between the experimental results of Tokanai et al. \cite{Tokanai2004} and developed Geant4 application simulation output with the ``G4LowEPPhysics" electromagnetic physics constructor. Of the different incident X-ray energy, scattering target, and metal filter combinations, the 40 keV incident X-ray energy, scattering target S2, and a Pb metal filter results (Simulation Set 3 in Fig. \ref{fig:r2} (bottom)) showed the largest difference between the normalised CdTe to YAP(Ce) detector signal ratios. In Fig. \ref{fig:r2} (bottom) the experimental results of Tokanai et al. \cite{Tokanai2004} normalised CdTe to YAP(Ce) detector signal ratios appear to have a high level of correlation with the simulated 625 \textmu m Pb metal filter results across the entire azimithal angle ($\phi$) range. The fact that experimental results of Tokanai et al. \cite{Tokanai2004} more closely aligns to the simulation results for a Pb metal filter that is 25 \textmu m thicker can be attributed that Pb metal sheets of less than 1 mm are very delicate, to the point that they can be easily torn or distorted through standard handling, and typically have manufacturing tolerance on the order of 50 to 100 \textmu m. Furthermore, this data represents a less than $\pm$ 5 \% difference from the true simulated metal filter thickness which is on the order of the uncertainty of Geant4's X-/gamma ray interaction cross-sectional data \cite{Cirrone2010,Arce2020}. 

\section{Conclusion}
\label{sec:C}

The Geant4 ``G4LowEPPhysics" electromagnetic physics constructor has been reconfigured to offer a ``best set" of electromagnetic physics models for studies exploring the transport of low energy polarised X-/gamma rays. An overview of the physics models implemented in ``G4LowEPPhysics", and it's experimental validation against Compton X-ray polarimetry measurements of the BL38B1 beamline at the SPring-8 synchrotron (Sayo, Japan) was reported through the use of a custom Geant4 application. It was found that ``G4LowEPPhysics" is able to reproduce the experimental results obtained at the BL38B1 beamline (SPring-8) to within a level of accuracy on the same order as Geant4's X-/gamma ray interaction cross-sectional data uncertainty (approximately $\pm$ 5 \%). This reconfigured version of ``G4LowEPPhysics" is intended to improve Geant4 user experience when undertaking polarised X-/gamma ray studies, and will be reviewed with each new Geant4 release from version 11.0.

\section*{Acknowledgements}

J.~M.~C.~Brown would like to acknowledge Professor Fuyuki Tokanai of Yamagata University (Japan) for their helpful clarification of the experimental setup and methodology implemented in Tokanai et al. \cite{Tokanai2004}.

\newpage

\begin{thebibliography}{00}


\bibitem{Paganin2006}
  Paganin~D.~M. \textit{Coherent X-ray optics}, Oxford University Press on Demand (2006).

\bibitem{Panaccione2009}
  Panaccione~G., Vobornik~I., Fujii~J., Krizmancic~D., Annese~E., Giovanelli~L., Maccherozzi~F., Salvador~F., De~Luisa~A., Benedetti~D. and Gruden~A. \textit{Advanced photoelectric effect experiment beamline at Elettra: A surface science laboratory coupled with Synchrotron Radiation}, Review of Scientific Instruments 80(4): 043105 (2009).

\bibitem{Strocov2010}
  Strocov~V.~N., Schmitt~T., Flechsig~U., Schmidt~T., Imhof~A., Chen~Q., Raabe~J., Betemps~R., Zimoch~D., Krempasky~J. and Wang~X. \textit{High-resolution soft X-ray beamline ADRESS at the Swiss Light Source for resonant inelastic X-ray scattering and angle-resolved photoelectron spectroscopies}, Journal of Synchrotron Radiation 17(5): 631-643 (2010).

\bibitem{Yamamoto2014}
  Yamamoto~S. et al. \textit{New soft X-ray beamline BL07LSU at SPring-8}, Journal of Synchrotron Radiation 21(2): 352-365 (2014).

\bibitem{Novik1975}
  Novick~R. \textit{Stellar and solar X-ray polarimetry}, Space Science Reviews 18(3): 389-408 (1975).

\bibitem{Meszaros1988}
  Meszaros~P., Novick~R., Szentgyorgyi~A., Chanan~G.~A. and Weisskopf~M~C. \textit{Astrophysical implications and observational prospects of X-ray polarimetry} The Astrophysical Journal 324: 1056-1067 (1988).

\bibitem{Lei1997}
  Lei~F., Dean~A.~J. and Hills~G.~L. \textit{Compton polarimetry in gamma-ray astronomy}, Space Science Reviews 82(3-4: 309-388 (1997).

\bibitem{Kumar2015}
  Kumar~P. and Zhang~B. \textit{The physics of gamma-ray bursts and relativistic jets}, Physics Reports 561: 1-109 (2015).

\bibitem{McConnell2017}
  McConnell~M.~L. \textit{High energy polarimetry of prompt GRB emission}, New Astronomy Reviews 76: 1-21 (2017).


\bibitem{Strocov2014}
  Strocov~V.~N., Wang~X., Shi~M., Kobayashi~M., Krempasky~J., Hess~C., Schmitt~T. and Patthey~L. \textit{Soft-X-ray ARPES facility at the ADRESS beamline of the SLS: Concepts, technical realisation and scientific applications}, Journal of Synchrotron Radiation 21(1): 32-44 (2014).
  
\bibitem{Kummer2016}
  Kummer~K., Fondacaro~A., Jimenez~E., Velez-Fort~E., Amorese~A., Aspbury~M., Yakhou-Harris~F., Van~Der~Linden~P. and Brookes~N.~B. \textit{The high-field magnet endstation for X-ray magnetic dichroism experiments at ESRF soft X-ray beamline ID32}, Journal of Synchrotron Radiation 23(2): 464-473 (2016).  
  
\bibitem{Hoesch2017}
  Hoesch~M. et al. \textit{A facility for the analysis of the electronic structures of solids and their surfaces by synchrotron radiation photoelectron spectroscopy}, Review of Scientific Instruments 88(1): 013106 (2017).

\bibitem{Schonfelder1993}
  Schonfelder~V. et al. \textit{Instrument description and performance of the imaging gamma-ray telescope COMPTEL aboard the Compton Gamma-Ray Observatory}, Astrophysical Journal Supplement Series (1993).

\bibitem{Lin2003}
  Lin~R.~P. et al. \textit{The Reuven Ramaty high-energy solar spectroscopic imager (RHESSI)},The Reuven Ramaty High-Energy Solar Spectroscopic Imager (RHESSI), Springer (2003).

\bibitem{Ubertini2003}
  Ubertini~P. et al. \textit{IBIS: The Imager on-board INTEGRAL}, Astronomy \& Astrophysics 411(1): L131-L139 (2003).

\bibitem{Vedrenne2003}
  Vedrenne~G. et al. \textit{SPI: The spectrometer aboard INTEGRAL}, Astronomy \& Astrophysics 411(1): L63-L70 (2003).

\bibitem{Kamae2008}
  Kamae~T. et al. \textit{PoGOLite–A high sensitivity balloon-borne soft gamma-ray polarimeter}, Astroparticle Physics 30(2): 72-84 (2008).

\bibitem{Singh2014}
  Singh~K.P. et al. \textit{ASTROSAT mission}, Space Telescopes and Instrumentation 2014: Ultraviolet to Gamma-Ray 9144: 91441S (2014).
  
\bibitem{Giomi2017}
  Giomi~M., B\"{u}hler~R., Sgr\'{o}, C., Longo~F., Atwood~W.~B. and Fermi LAT Collaboration, \textit{Estimate of the Fermi large area telescope sensitivity to gamma-ray polarization}, AIP Conference Proceedings 1792(1): 070022 (2017).

\bibitem{Zoglauer2006}
  Zoglauer~A., Andritschke~R. and Schopper~F. \textit{MEGAlib–the medium energy gamma-ray astronomy library}, New Astronomy Reviews 50(7-8): 629-632 (2006).

\bibitem{Agapov2009}
  Agapov~I., Blair~G.~A., Malton~S. and Deacon~L. \textit{BDSIM: A particle tracking code for accelerator beam-line simulations including particle–matter interactions}, Nuclear Instruments and Methods in Physics Research Section A 606(3): 708-712 (2009).

\bibitem{Bulgarelli2012}
  Bulgarelli~A., Fioretti~V., Malaguti~P., Trifoglio~M. and Gianotti~F. \textit{BoGEMMS: the Bologna Geant4 multi-mission simulator},High Energy, Optical, and Infrared Detectors for Astronomy 8453: 845335 (2012).

\bibitem{Chattopadhyay2013}
  Chattopadhyay~T., Vadawale~S.~V. and Pendharkar~J. \textit{Compton polarimeter as a focal plane detector for hard X-ray telescope: sensitivity estimation with Geant4 simulations}, Experimental Astronomy 35(3): 391-412 (2013).

\bibitem{Cornelius2014}
  Cornelius~I., Guatelli~S., Fournier~P., Crosbie~J.~C., Sanchez~del~Rio~M., Br\"{a}uer-Krisch~E., Rosenfeld~A. and Lerch~M. \textit{Benchmarking and validation of a Geant4–SHADOW Monte Carlo simulation for dose calculations in microbeam radiation therapy}, Journal of Synchrotron Radiation 21(3): 518-528 (2014).

\bibitem{Dimmock2015}
  Dimmock~M.~R., de~Jonge~M.~D., Howard~D.~L., James~S.~A., Kirkham~R., Paganin~D.~M., Paterson~D.~J., Ruben~G., Ryan~C.~G. and Brown~J.~M.~C. \textit{Validation of a Geant4 model of the X-ray fluorescence microprobe at the Australian synchrotron}, Journal of Synchrotron Radiation 22(2): 354-365 (2015).

\bibitem{Moiseev2015}
  Moiseev~A.~A. et al. \textit{Compton-pair production space telescope (ComPair) for MeV gamma-ray astronomy}, ArXiv 1508.07349 (2015).

\bibitem{Chauvin2016}
  Chauvin~M., Jackson~M., Kawano~T., Kiss~M., Kole~M., Mikhalev~V., Moretti~E., Takahashi~H. and Pearce~M. \textit{Optimising a balloon-borne polarimeter in the hard X-ray domain: From the PoGOLite Pathfinder to PoGO+}, Astroparticle Physics 82: 99-107 (2016).

\bibitem{G42003}
  Agostinelli~S. et al. \textit{Geant4-a simulation toolkit}, Nuclear Instruments and Methods in Physics Research Section A 506(3): 250-303 (2003).

\bibitem{G42006}
  Allison~J. et al. \textit{Geant4 developments and applications}, IEEE Transactions on Nuclear Science 53(1): 270-278 (2006).

\bibitem{G42016}
  Allison~J. et al. \textit{Recent developments in Geant4}, Nuclear Instruments and Methods in Physics Research Section A 835: 186-225 (2016).

\bibitem{G4Phys2020}
  GEANT4 Collaboration, \textit{GEANT4 Physics Reference Manual - Version 10.6},  https://geant4.web.cern.ch (2020).

\bibitem{G41994}
  Dell’Acqua~A. et al. \textit{Geant 4: an object-orientated toolkit for simulation in HEP}, Technical Report DRDC-94-29, CERN (1994).


\bibitem{Schalick2007}
  Sch\:{a}licke~A., Laihem~K. and Starovoitov~P. \textit{Polarised Geant4-Applications at the ILC} ArXiv 0712.2336 (2007).
  
\bibitem{Depaola2009}
  Depaola~G.~O. and Iparraguirre~M.~L. \textit{Angular distribution for the electron recoil in pair production by linearly polarized $\gamma$-rays on electrons}, Nuclear Instruments and Methods in Physics Research Section A 611(1): 84-92 (2009).
  
\bibitem{Bernard2018}
  Bernard~D. \textit{A 5D, polarised, Bethe–Heitler event generator for $\gamma$ to e+ e- conversion}, Nuclear Instruments and Methods in Physics Research Section A 899: 85-93 (2018).

\bibitem{Semeniouk2019}
  Semeniouk~I. and Bernard~D. \textit{C++ implementation of Bethe–Heitler, 5D, polarized, Gamma to e+ e- pair conversion event generator}, Nuclear Instruments and Methods in Physics Research Section A 936: 290-291 (2019).

\bibitem{Omer2019}
  Omer~M. and Hajima~R. \textit{Validating polarization effects in $\gamma$-rays elastic scattering by Monte Carlo simulation}, New Journal of Physics 21(11): 113006 (2019).
 and
\bibitem{Gavrila1959}
  Gavrila~M. \textit{Relativistic K-shell photoeffect}, Physical Review 113(2): 514-526 (1959).

\bibitem{Gavrila1961}
  Gavrila~M. \textit{Relativistic L-shell photoeffect}, Physical Review 124(4): 1132-1141 (1961).

\bibitem{Depaola2003}
  Depaola~G.~O. \textit{New Monte Carlo method for Compton and Rayleigh scattering by polarized gamma rays}, Nuclear Instruments and Methods in Physics Research Section A 512(3): 619-630 (2003).

\bibitem{Brown2014}
  Brown~J.~M.~C., Dimmock~M.~R., Gillam~J.~E. and Paganin~D.~M. \textit{A low energy bound atomic electron Compton scattering model for Geant4}, Nuclear Instruments and Methods in Physics Research Section B 338: 77-88 (2014).

\bibitem{Seltzer1989}
  Seltzer~S.~M., Perkins~S.~T. and Cullen~D.~E. \textit{Tables and graphs of electron-interaction cross-sections from 10 eV to 100 GeV derived from the LLNL evaluated electron data library (EEDL): Z=1-100}, Technical Report UCRL-50400 Vol.31, Lawrence Livermore National Laboratory (1989).

\bibitem{LassilaPerin1995}
  Lassila-Perini~K. and Urb\'{a}n~L. \textit{Energy loss in thin layers in GEANT}, Nuclear Instruments and Methods in Physics Research Section A: 362(2-3): 416-422 (1995).

\bibitem{Allison2012}
  Allison~J. et~al. \textit{Geant4 electromagnetic physics for high statistic simulation of LHC experiments}, Journal of Physics - Conference Series 396(2): 022013 (2012).

\bibitem{Bagulya2017}
  Bagulya~A. et al. \textit{Recent progress of GEANT4 electromagnetic physics for LHC and other applications},Journal of Physics - Conference Series 898: 042032 (2017).  
  
\bibitem{Ivanchenko2010}
  Ivanchenko~V.~N., Kadri~O., Maire~M. and Urban~L. \textit{Geant4 models for simulation of multiple scattering}, Journal of Physics - Conference Series 219: 032045 (2010).
 
\bibitem{Heitler1954} 
  Heitler~W. \textit{The Quantum Theory of Radiation}, Oxford Clarendon Press (1954).
  
\bibitem{Nelson1985} 
  Nelson~W.~R., Hirayama~H. and Rogers~D.~W. \textit{EGS4 code system} (No. SLAC-265), Stanford Linear Accelerator: SLAC-265 (1985).
  
\bibitem{Salvat2008}
  Salvat~F., Fern\'{a}ndez-Varea~J.~M. and Sempau~J. \textit{PENELOPE-2008: A code system for Monte Carlo simulation of electron and photon transport}, Workshop Proceedings Barcelona (Spain), OECD/NEA (2008).

\bibitem{Tokanai2004}
  Tokanai~F., Sakurai~H., Gunji~S., Motegi~S., Toyokawa~H., Suzuki~M., Hirota~K., Kishimoto~S. and Hayashida~K. \textit{Hard X-ray polarization measured with a Compton polarimeter at synchrotron radiation facility}, Nuclear Instruments and Methods in Physics Research Section A: 530(3): 446-452 (2004).
  
\bibitem{Redus2007}
  Redus, R. \textit{Charge trapping in XR-100T-CdTe and-CZT detectors}, Application Note ANCZT-2 Rev. 3 (2007).
  
\bibitem{Cirrone2010}
  Cirrone~G.~A.~P., Cuttone~G., Di~Rosa~F., Pandola~L., Romano~F. and Zhang~Q. \textit{Validation of the Geant4 electromagnetic photon cross-sections for elements and compounds}, Nuclear Instruments and Methods in Physics Research Section A 618(1-3): 315-322 (2010).
  
\bibitem{Arce2020}
  Arce~P. et al. \textit{Report on G4‐Med: a Geant4 benchmarking system for medical physics applications developed by the Geant4 Medical Simulation Benchmarking Group}, Medical Physics: In Press (2020).
  
\end{thebibliography}
\end{document}